\documentclass{article}
\usepackage{spconf,amsmath,graphicx}
\usepackage[ruled,vlined]{algorithm2e}
\usepackage{amsfonts} 
\usepackage{booktabs}
\usepackage{float}
\usepackage{multirow}
\usepackage{subcaption}

\title{Blind signal dereverberation for machine speech recognition}
\name{Samik Sadhu$^1$, Hynek Hermansky$^{1,2}$}
\address{ $^1$Center for Language and Speech Processing, Johns Hopkins University, USA\\
  $^2$Human Language Technology Center of Excellence, Johns Hopkins University, USA}
\begin{document}
%
\maketitle
\begin{abstract}
We present a method to remove unknown convolutive noise introduced to speech by reverberations of recording environments, utilizing some amount of training speech data from the reverberant environment, and any available non-reverberant speech data.  Using  Fourier transform computed over long temporal windows, which ideally cover the entire room impulse response,  we convert room induced convolution to additions in the log spectral domain. Next, we compute a spectral normalization vector from statistics gathered over reverberated as well as over clean speech in the log spectral domain. During operation, this normalization vectors are used to alleviate reverberations  from complex speech spectra recorded under the same reverberant conditions . Such dereverberated complex speech spectra are used to compute complex FDLP-spectrograms for use in automatic speech recognition.
\end{abstract}
\section{Introduction}
\label{sec:intro}

In many speech-to-text (STT) applications, the message-carrying speech signal $s(t)$ is corrupted by room reverberations $n(t)$, yielding the reverberated signal $o(t)=s(t) \ast n(t)$ where $\ast$ is the convolution operator and $t$ denotes time. According to the convolution theorem of Fourier transform, convolutions in time domain, turn into multiplication in the spectral domain and thereby additions in log spectral domain shown in equation \ref{eq:main_idea}.
\begin{eqnarray}
\label{eq:main_idea}
    \log \mathcal{F}(o(t)) &=&  \log \mathcal{F}(o(t)) \\ \nonumber 
          &=& \log \mathcal{F}(s(t) \ast n(t))  \\ \nonumber 
    &=& \log ( \mathcal{F}(s(t)) \times \mathcal{F}(n(t))) \\ \nonumber 
    &=&\log \mathcal{F}(s(t)) + \log  \mathcal{F}(n(t)) 
\end{eqnarray}
$\mathcal{F}$ indicates the Fourier transform operator. Thus, for known $n(t)$, the original signal $s(t)$ could easily be recovered as 
\begin{equation}
\label{eq:main_idea2}
    s(t)=\mathcal{F}^{-1} ( \exp (\log  \mathcal{F}(o(t)) -\log  \mathcal{F}(n(t))) )
\end{equation}, 
where $\mathcal{F}^{-1}$ is the inverse Fourier transform operator. 

However, a few practical issues arise in this analysis. 
\begin{itemize}
    \item Even though $n(t)$ has infinite duration, in digital signal processing, it can only be represented as a finite length discrete time signal.
    \item In order to use equation \ref{eq:main_idea2} with digital signals, we need equal length discrete Fourier transforms of the digitized versions of the signals $o(t)$ and $n(t)$.
    \item As we shall show, arithmetic operations done in the log spectral domain needs phase unwrapping operations to remove phase ambiguity.
    \item $n(t)$ is typically not known.
\end{itemize}

\section{Proposed Technique}
Suppose that the infinitely long impulse response $n(t)$ can be approximated by its truncated digital version $n^{\prime}=\{n^{\prime}_k\}_{k=1}^S$. Even though $n^{\prime}$ is unknown, it can be assumed to be a constant vector of real numbers. 

Assume a digitized observed reverberated speech utterances $o=\{o_k\}_{k=1}^{S+T-1}$, obtained by convolving a source signal $s=\{s_k\}_{k=1}^T$ with $n^{\prime}$. Appropriate number of zeros can be appended to the signals to make them uniform length sequences leading to the equation 
\begin{equation}
\label{eq:discrete_main}
    \log \mathbb{F}(o) =\log \mathbb{F}(s) + \log \mathbb{F}(n^{\prime})
\end{equation}, 
where $\mathbb{F}$ is the discrete Fourier transform operator. Since $n^{\prime}$ is assumed to be a constant vector, so is $\log \mathbb{F}(n^{\prime})$. Computing the expected values on both sides of equation \ref{eq:discrete_main}, we get 
\begin{equation}
\label{eq:discrete_main_extecpted}
    \mathbb{E}\log \mathbb{F}(o) =\mathbb{E} \log \mathbb{F}(s) + \log \mathbb{F}(n^{\prime})
\end{equation}
Thus, an estimate of the unchanging logarithmic spectra of the room impulse response can simply be obtained as 
\begin{equation}
\label{eq:get_rir}
    \log \mathbb{F}(n^{\prime}) = \mathbb{E}\log \mathbb{F}(o) -\mathbb{E} \log \mathbb{F}(s) 
\end{equation}
Equation \ref{eq:get_rir} forms the basis of our algorithm where the expected values are replaced with \textit{empirical sums} computed over a finite number of speech utterances to obtain an estimate $\phi$ of the log spectrum of the room impulse response $\log \mathbb{F}(n^{\prime})$. This estimate can be used to normalize the log spectrum of the observed speech to estimate the clean speech as

\begin{equation}
\label{eq:main_idea2_discrete}
    \hat{s}=\mathbb{F}^{-1} ( \exp (\log  \mathbb{F}(o) - \phi ) )
\end{equation} 

\subsection{Relation to prior work}
This principle follows Stockham et al. \cite{1451730} with several important differences, which come from our application to dereverberation. Stockham et al. aimed for mitigating effects of distortions, which resulted from use of a primitive recording machine. Impulse response of such a machine was relatively short comparing to our longer response of the reverberant room. Therefore, unlike Stockham  et al. , who were dealing with deconvolving rather short impulse responses, we are dealing with much longer impulse responses of rooms. Subsequently, our applied time segments are significantly scaled up. While they worked with 500 millisecond segments and averaged over several minutes of the recording on the old-fashioned gramophone plate, we used speech segments representing whole utterances and use as much as 180 hours of reverberant speech for computing the averages. Dealing with large room impulse responses also forces us to work with spectral subtractions in complex domain, while Stockham  et al. worked with magnitude spectra and used phases which were computed from recovered magnitude spectra under minimum-phase assumption. This is unsatisfactory in our application. 

\subsection{The algorithm}
Our algorithm is requires the availability of two sets of speech data.
\begin{itemize}
    \item \textbf{Data1}: $P_o$ utterances $o^{(i)}$, $i=1,2,\dots P_o$ from the domain of interest which generally will be a reverberant environment. For the baseline experiment, this can also be clean speech data as we explain in section \ref{sec:exp}.\\
    \item \textbf{Data2}: $P_s$ utterances $s^{(i)}=1,2,\dots P_s$ of high quality clean speech data. This data does not need to be related to \textbf{Data1}.
\end{itemize}
Additionally, the room impulse response $n^{\prime}$ is assumed to be a fixed vector for all $P_o$ utterances in \textbf{Data1}. 
\subsubsection{Computing the normalization vector $\phi$}
We compute $\phi$ following three simple steps.
\begin{itemize}
    \item Append zeros to each utterance from \textbf{Data1} and \textbf{Data2} to make all of them have uniform number of samples $N_{uniform}$.  
    \item Compute Fourier transforms of each zero appended utterance $\mathbb{F}( o^{(i)} )$, $i=1,2,\dots P_o$ and $\mathbb{F}( s^{(i)} )$, $i=1,2,\dots P_s$ 
    \item Compute the normalizing vector \\
    $\phi=\frac{1}{P_o} \sum_{i=1}^{P_o}\log \mathbb{F}( o^{(i)} ) - \frac{1}{P_s}  \sum_{i=1}^{P_s} \log \mathbb{F}( s^{(i)} )$
\end{itemize}
During operation $\phi$ can be used to dereverberate speech utterances. 
\subsubsection{Utterance-wise log spectrum normalization}
\label{sec:normalized_speech}
\begin{enumerate}
    \item Consider an observed test utterance $o^{test}$. Append zeros to make it $N_{uniform}$ samples long
    \item Compute Fourier transform $\mathbb{F} (o^{test})$
    \item Compute normalized speech \\
    $\hat{s}^{test}=\mathbb{F}^{-1} (\exp (\log \mathbb{F} (o^{test})-\phi)$ )
\end{enumerate}
This normalized speech utterance can be used for extract features for speech recognition. However, this leads to a processing delay amounting to the duration of each utterance. Alternately, we can directly compute FDLP-spectrograms from 1.5 seconds long windowed speech leading to lesser delays. 

\subsubsection{Frame-wise log spectrum normalization and direct FDLP-spectrogram feature extraction }
\label{sec:asr_feats}
\begin{enumerate}
    \item For an observed test utterance $o^{test}$, generate von Hann windowed frames covering samples over 1.5 second of speech with 50\% overlap. 
    \item Considering there are $M$ of these frames $o^{test,m}$, $m=1,2,\dots M$, append each frame with zeros to make them $N_{uniform}$ samples long. 
    \item Compute Fourier transform of each frames $\mathbb{F} ( o^{test,m} )$, $m=1,2,\dots M$
    \item Compute normalized speech frames \\
    $\hat{o}^{test,m}=\mathbb{F}^{-1} (\exp (\log \mathbb{F}(o^{test,m})-\phi)$ )
    \item Use $\hat{o}^{test,m}$, $m=1,2,\dots M$ to compute complex FDLP-spectrogram\cite{sadhu2022complex}.
\end{enumerate}
Elegantly combining feature extraction with dereverberation, frame-wise normalization requires the analysis window to be significantly larger than the length of the room impulse response. 
\subsubsection{Phase wrapping issues for computing $\phi$}
For any $i \in \{ 1,2, \dots P_o\}$, the discrete Fourier transform $\mathbb{F} (o^{(i)})$ written in exponential form is given by \\ $\mathbb{F} (o^{(i)})=|\mathbb{F} (o^{(i)})| \exp ( j \angle \mathbb{F} (o^{(i)}) )$, where $|*|$ and $\angle *$ indicates the standard magnitude and phase operators on any complex number respectively. In the log domain, this exponential form becomes 
\begin{equation}
    \log \mathbb{F} (o^{(i)})= \log |\mathbb{F} (o^{(i)})| + j \angle \mathbb{F} (o^{(i)})
\end{equation}
and hence
\begin{equation}
\label{eq:the_issue}
    \frac{1}{P_o} \sum_{i=1}^{P_o}\log \mathbb{F}( o^{(i)} ) = \frac{1}{P_o} \sum_{i=1}^{P_o}\log |\mathbb{F}( o^{(i)} )| + j \frac{1}{P_o} \sum_{i=1}^{P_o} \angle \mathbb{F}( o^{(i)} )
\end{equation}.
 Although computing the first term on the right hand side of equation \ref{eq:the_issue} is straight forward, the second term is highly ambiguous because of phase wrapping. Since phase is an angular variable, it takes on values between 0 to $2 \pi$. Phase angles beyond that are reduced to this range by adding or subtracting appropriate multiples of $2\pi$ from it. For example, two complex numbers $1\angle j\pi$ and $1\angle j3\pi$ are identical. However, averaging the phases of $1\angle j2\pi$ and $1\angle j\pi$ gives us $1.5\pi$, whereas average phase of $1\angle j2\pi$ and $1\angle j3\pi$ is $2.5\pi$. A detailed description of this phase wrapping problem can be found in \cite{oppenheim1968nonlinear}.
 
 To disambiguate the second term in eq. \ref{eq:the_issue}, we unwrap the phases \cite{schafer1969echo} of \textit{all} computed spectra using the following simple algorithm. Consider a wrapped discrete phase sequence $p=\{p_k\}_{k=1}^N$ where $0 \leq p_k \leq 2\pi$ $\forall k \in \{1,2,\dots N\}$. To compute the unwrapped phase $\hat{p}$, we follow algorithm \ref{algo:1}.

\begin{algorithm}
\caption{Phase Unwrapping $p \xrightarrow[]{} \hat{p} $} \label{algo:1}
Initialize unwrapped phase $\hat{p}_k=0$ $\forall k \in \{1,2,\dots N\}$ \;
 $\hat{p}_1=p_1$ \;
\For{$k=2 \to N$}
{
$\hat{p}_k=p_k+2\pi t$ \;
$t$ \text{chosen s.t.} $|\hat{p}_k- \hat{p}_{k-1}| \leq \pi$  \;
}
\end{algorithm}

\section{Experimental Details}
\label{sec:exp}
\subsection{Data sets}
\subsubsection{Training data}
For our experiments we obtain the following \textit{mutually exclusive} segments from the Librispeech \cite{librispeech} data. 
\begin{itemize}
    \item \textbf{High quality clean speech data set}: 180 hours of clean speech that is used as \textbf{Data2} for all our experiments.
    \item \textbf{Four different reverberated data sets}: 180 hours of clean speech data that is convolved with four different room impulse responses to generate four separate reverberated speech data sets. Each of them are used as \textbf{Data1} for computing the normalization vector $\phi$ when testing the ASR performance under a specific reverberant condition. 
    \item \textbf{ASR training data set}: 50 hours of clean speech data that is used for ASR training. In order to keep the speech features consistent across our experiments, we also normalize utterance-wise log spectrum of this data with $\phi_{clean}$, computed by using this 50 hours of clean data as \textbf{Data1}. 
\end{itemize}

\subsubsection{Room impulse responses}
In our experiments, four different types of room impulse responses provided with the REVERB data set are used, two of which are large room impulse responses with reverberation time $T_{60}=0.7$ seconds and another two medium sized room impulse responses with reverberation time $T_{60}=0.5$ seconds. $T_{60}$ is the time taken for the impulse response power to decay by 60dB. In all four cases, the source is located 250 cm away from the microphone at two different angles. 

\subsubsection{Test data}
To evaluate the performance of the ASR under clean test condition, we use the original clean Librispeech test set. Additionally, this clean test set is convolved with the same four room impulse responses in use to obtain respective test sets under each of the four reverberant conditions. $\phi_{clean}$ is used as the ``default" normalization vector under various testing conditions to obtain baseline ASR results on various test sets. 

\subsection{ASR setup}

\subsubsection{Front-end}

Our front-end uses FDLP-spectrogram features \cite{sadhu2021radically} using complex FDLP \cite{sadhu2022complex} to extract modulations from raw speech waveform. Speech utterances are segmented with 1.5 seconds long analysis windows with 50\% overlap.

\subsubsection{ASR architecture}
In our experiments, a Connectionist Temporal Classification (CTC) ASR acoustic model is used with 12 Transformer layers together along with a neural language model with 6 Transformer layers all trained with ESPnet \cite{watanabe2018espnet} speech recognition toolkit. 
\subsection{Generalized weighted prediction error based dereverberation}
Weighted Prediction Error (WPE) \cite{yoshioka2010speech} based blind dereverberation works by estimating the theoretically obtainable signal from only the first $\Delta$ taps of the room impulse response and optimizing for a prediction filter that eliminates the late reverberation effects. An improved optimization loss is called for in the more widely used Generalized WPE \cite{wpe} which is used in Section \ref{sec:results} for comparing recognition results with regards to our technique.

\section{Results}
\label{sec:results}
\begin{figure}[H]

\centering
\begin{subfigure}[b]{0.5\textwidth}
   \includegraphics[width=1\linewidth]{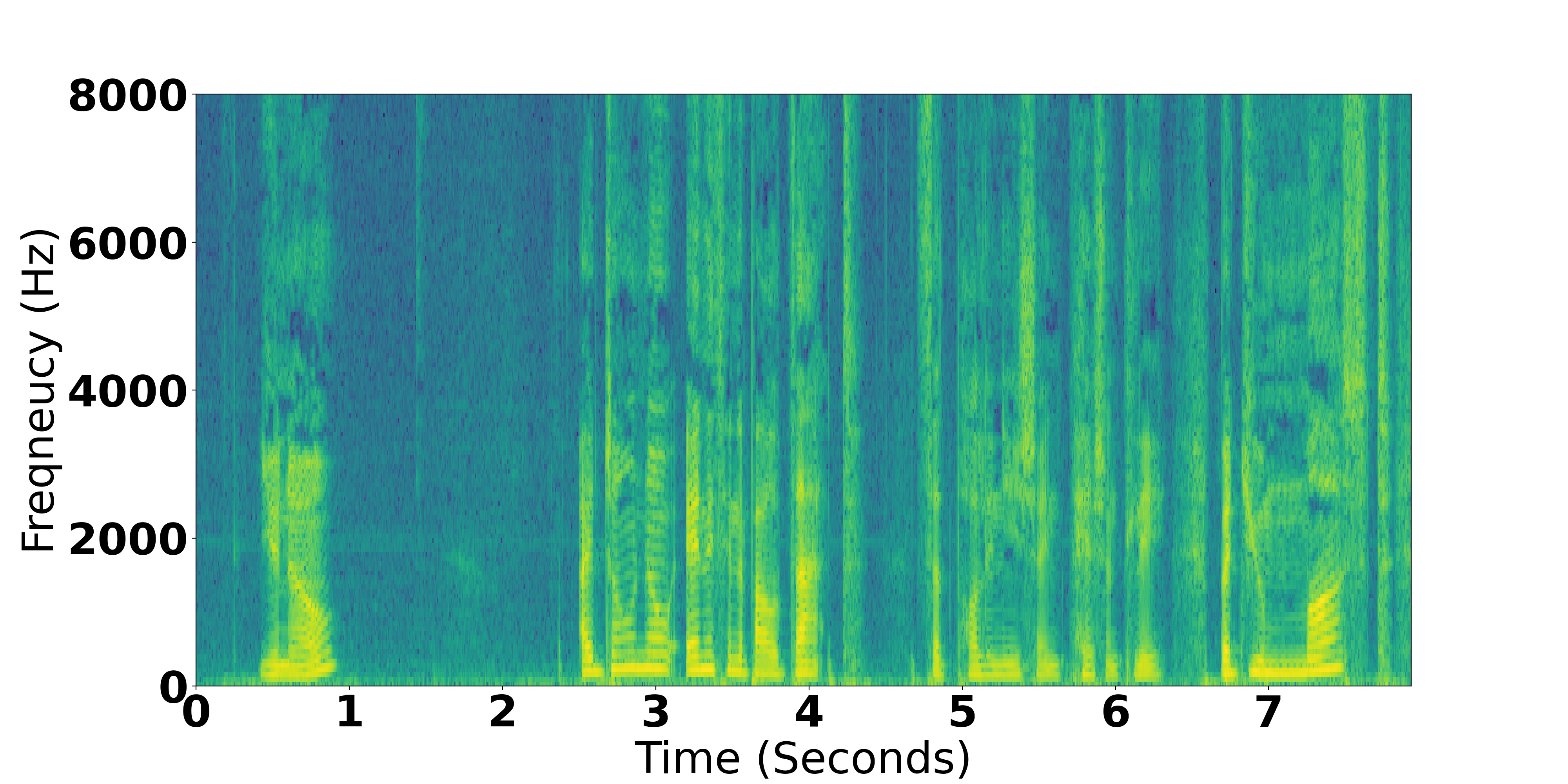}
   \caption{Original clean speech spectrogram}
   \label{fig:Ng1} 
\end{subfigure}

\begin{subfigure}[b]{0.5\textwidth}
   \includegraphics[width=1\linewidth]{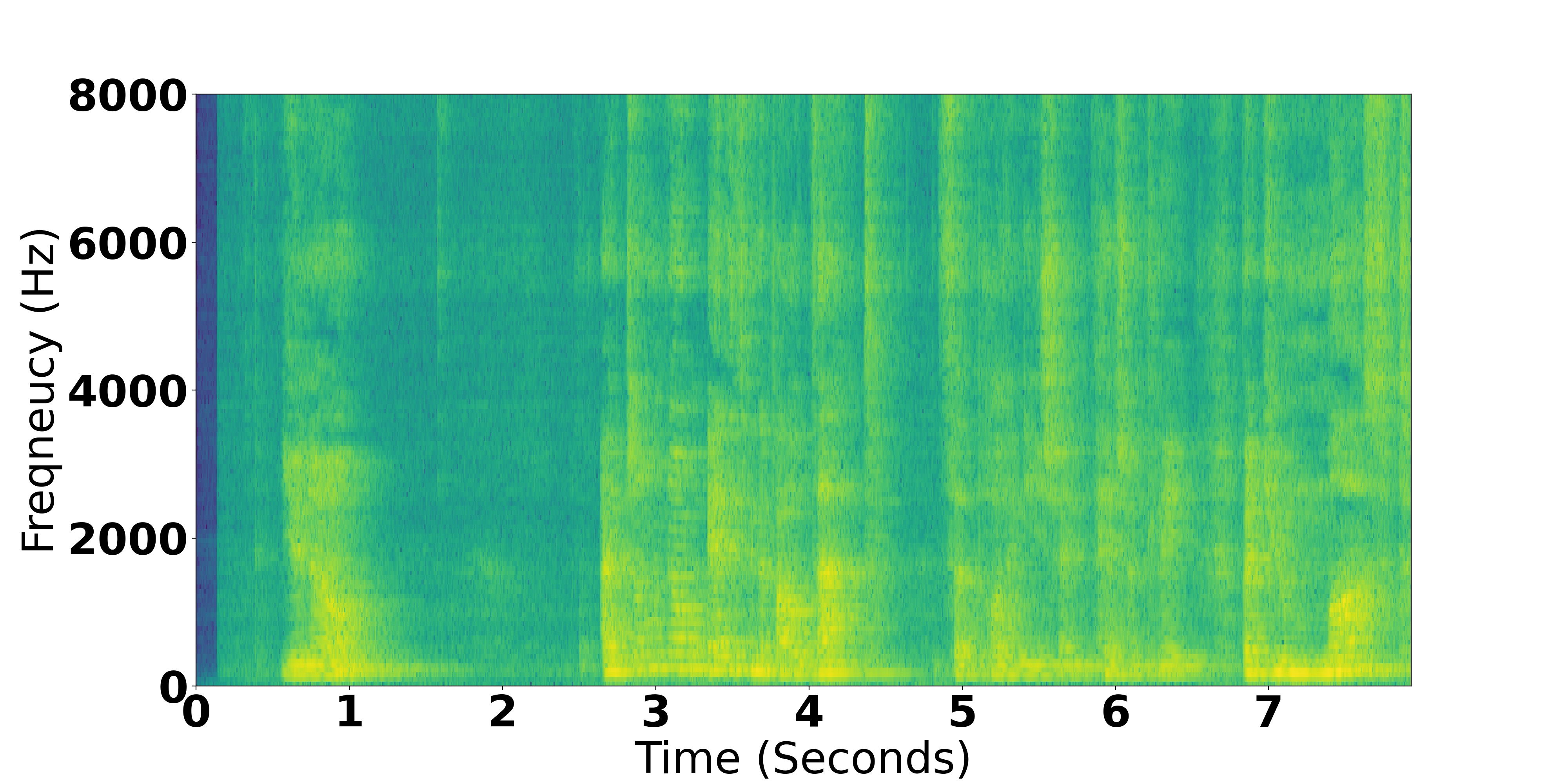}
   \caption{Reverberated speech spectrogram}
   \label{fig:Ng1} 
\end{subfigure}

\begin{subfigure}[b]{0.50\textwidth}
   \includegraphics[width=1\linewidth]{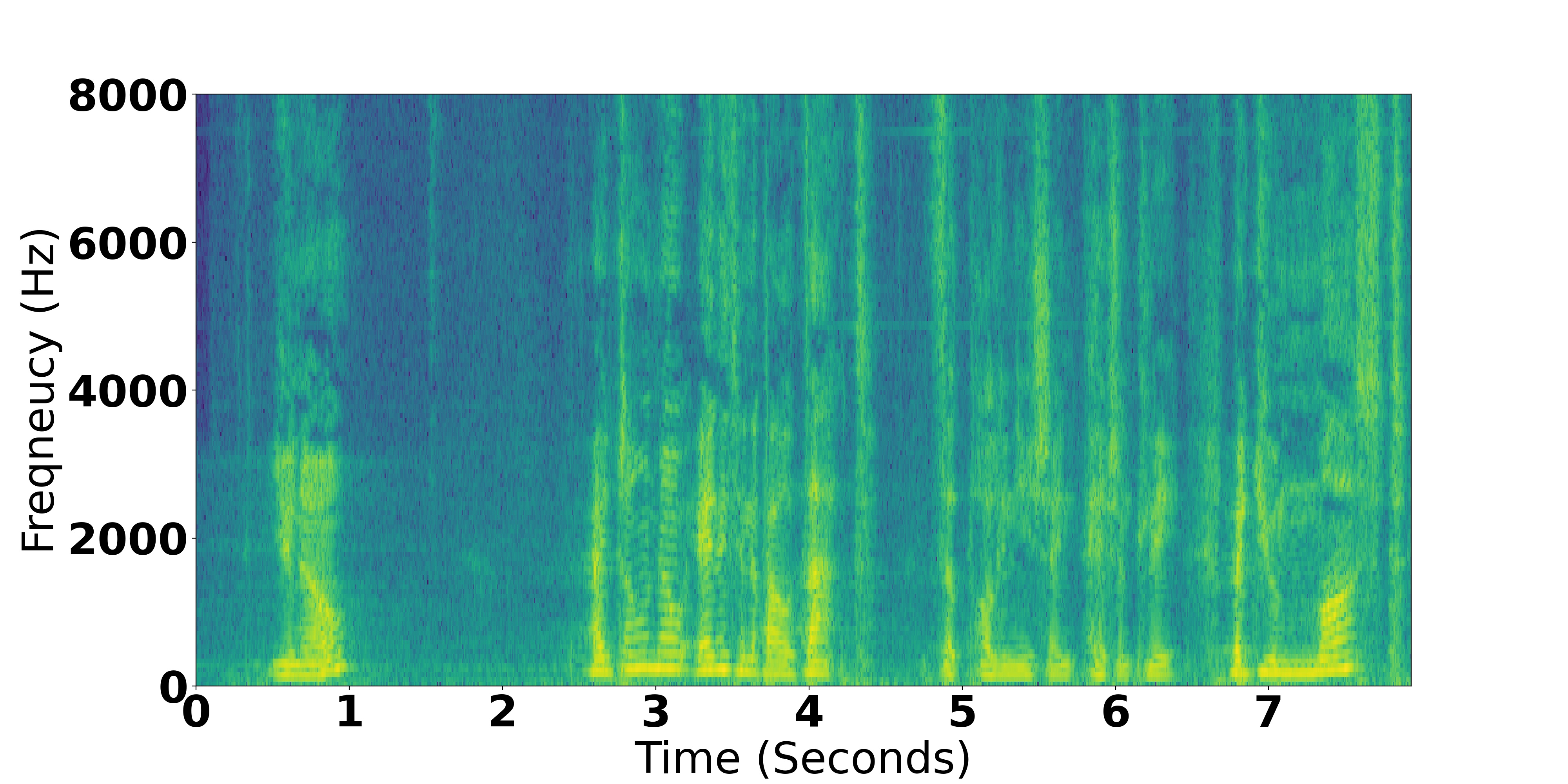}
   \caption{Spectrogram of same utterance after dereverberation}
   \label{fig:Ng2}
\end{subfigure}

\caption[x]{Result of dereverberating large room reverberated speech using the utterance-wise normalization described in Section \ref{sec:normalized_speech} }.
\label{fig:comapre_clean_and_noisy_spectrogram}
\end{figure}

Figure \ref{fig:comapre_clean_and_noisy_spectrogram} shows an example large room reverberant speech utterance and the dereverberation result when its log spectrum is utterance-wise normalized (as per Section \ref{sec:normalized_speech}).  The CTC ASR when tested on clean librispeech test set using utterance-wise $\phi_{clean}$ normalization yields a Word Error Rate (WER) of 26.2\%. This provides a lower bound on ASR performance under any of the reverberant conditions. 

\subsection{Recognition results with utterance-wise normalization}
Table \ref{table:recognition_comparison_uttwise} compares the results of three different speech recognition pipelines under various reverberant test conditions. Not surprisingly, using our ASR as an off-the-shelf speech recognizer over the four reverberant conditions yields significantly inferior recognition results. Recognition accuracy is universally benefited to a certain degree by the addition of generalized WPE dereverberation front-end to the ASR. However, the improvements offered by our method is remarkable and closest to the clean test set error rate of 26.2\%.

\begin{table}[h]
\centering
\begin{tabular}{@{}lccc@{}}
\toprule 
\multirow{2}{*}{\begin{tabular}[c]{@{}l@{}}reverberant\\ condition\end{tabular}} & \begin{tabular}[c]{@{}c@{}}default \\ normalization\end{tabular} & \begin{tabular}[c]{@{}c@{}}default\\ normalization \\ +WPE\end{tabular} & \begin{tabular}[c]{@{}c@{}}our  \\ method\end{tabular} \\ \cmidrule(l){2-4} 
                                                                                 & \multicolumn{3}{c}{WER \%}                                                                                                                                                                                         \\ \midrule
\multicolumn{1}{l|}{large room 1}                                                & 75.7                                                                    & 70.2                                                                            & 29.4                                                   \\
\multicolumn{1}{l|}{large room 2}                                                & 80.2                                                                    & 78.4                                                                            & 30.2                                                   \\
\multicolumn{1}{l|}{medium room 1}                                               & 62.8                                                                    & 56.3                                                                            & 29.2                                                   \\
\multicolumn{1}{l|}{medium room 2}     &    76.3            &       70.9            &   29.3                                                     \\ \bottomrule
\end{tabular}
\caption{ASR test results under four reverberant conditions are shown. The \textbf{first column} of results shows the recognition result with no extra front-end processing except utterance-wise speech log spectrum normalization by $\phi_{clean}$. The \textbf{second column} shows the results with WPE front-end dereverberation of the raw waveform before $\phi_{clean}$ normalization. The \textbf{third column} shows the recognition result of obtaining normalized speech utterances using the appropriate normalization vector as described in Section \ref{sec:normalized_speech}.}
\label{table:recognition_comparison_uttwise}
\end{table}

\subsection{Recognition results with frame-wise normalization}
Table \ref{table:recognition_comparison_phramewise} shows that speech recognition with frame-wise normalization works slightly inferior to utterance-wise normalization. When using windowed speech frames, our foundational idea expressed in equation \ref{eq:discrete_main} approximately holds true when the window length is considerably larger than the length of the room impulse response \cite{1451730}. In our case, a compromise is reached with an analysis window of 1.5 seconds. This is twice the $T_{60}$ value of the longest room impulse response at hand and also allows for fast FDLP-spectrogram feature extraction with appropriate resolution in modulation frequency domain. This explains the reduction is recognition accuracy, although still significantly better than WPE dereverberation seen in table \ref{table:recognition_comparison_uttwise}.

\begin{table}[h]
\centering
\begin{tabular}{@{}lc@{}}
\toprule
\multirow{2}{*}{\begin{tabular}[c]{@{}l@{}}reverberant\\ condition\end{tabular}} & \begin{tabular}[c]{@{}c@{}}recognition using FDLP-spectrograms  \\  with 1.5 second long  \\ normalized log spectrum \end{tabular} \\ \cmidrule(l){2-2} 
                                                                                 & WER \%                                                                                                 \\ \midrule
\multicolumn{1}{l|}{large room 1}                                                & 37.5                                                                                                   \\
\multicolumn{1}{l|}{large room 2}                                                & 39.1                                                                                                   \\
\multicolumn{1}{l|}{medium room 1}                                               & 37.2                                                                                                   \\
\multicolumn{1}{l|}{medium room 2}                                               &  43.9                                                                                                      \\ \bottomrule
\end{tabular}
\caption{ASR test results under four reverberant conditions using normalized log spectrum of 1.5 seconds long windowed speech directly used for computing FDLP-spectrogram. (see  Section \ref{sec:asr_feats})}
\label{table:recognition_comparison_phramewise}
\end{table}

\section{Discussions and conclusions}
A straightforward method to remove stationary reverberations from speech signal is described in this paper. Our method utilizes fundamental concepts from convolution theorem of Fourier transform and empirically computed expectations of log spectrum to obtain a normalizer that can remove effect of reverberations from whole utterances as well as speech frames.

\begin{table}[H]
\centering
\begin{tabular}{@{}lc@{}}
\toprule
\multirow{2}{*}{\begin{tabular}[c]{@{}l@{}}amount of \\ reverberated \\ data used\end{tabular}} & \begin{tabular}[c]{@{}c@{}}recognition under \\ large room reverberant \\ condition\end{tabular} \\ \cmidrule(l){2-2} 
                                                                                                & WER \%                                                                                              \\ \midrule
\multicolumn{1}{l|}{50 hours}                                                                   & 42.9                                                                                                \\
\multicolumn{1}{l|}{100 hours}                                                                  &   31.9                                                                                                  \\ \bottomrule
\end{tabular}
\caption{ASR recognition results under a large room reverberant condition with variable amounts of reverberant data (\textbf{Data1}) used for computing the normalization vector. }
\label{table:amount_of_data}
\end{table}

Dereveberation does not require any labelled training data. For the unlabeled data requirements, the clean speech data is abundant. However, reverberant data collected from a fixed microphone setup from the same room, could be more difficult to collect. In the idealized laboratory situation presented in this paper, the signal was artificially reverberated so the convolving system was fixed for each database whereas in practical applications, the reverberations may be changing their character. To get the average logarithmic spectrum, 180 hours of reverberant data was used which might not be available from one fixed reverberant condition. Table \ref{table:amount_of_data} shows the degradation in recognition performance for reduced amount of reverberant data. It is apparent that our method depends on significant amount of reverberant data to be effective. This is clearly a current drawback of our technique for most practical situations and further efforts are under way to better understand this aspect. 

 Resulting FDLP-spectrogram features are shown to be exceptionally robust to the effects of reverberation for speech recognition.

\section{ACKNOWLEDGMENTS}
\label{sec:ack}
This work was supported  by the Center of Excellence in Human Language Technologies, The Johns Hopkins University.

\bibliographystyle{IEEEbib}
\bibliography{strings,refs}

\end{document}